# Feature importance recap and stacking models for forex price prediction

Yunze Li, Yanan Xie, Chen Yu , Fangxing Yu, Bo Jiang and Matloob Khushi

The School of Computer Science, The University of Sydney

## ABSTRACT

Forex trading is the largest market in terms of qutantitative trading. Traditionally, traders refer to technical analysis based on the historical data to make decisions and trade. With the development of artificial intelligent, deep learning plays a more and more important role in forex forecasting. How to use deep learning models to predict future price is the primary purpose of most researchers. Such prediction not only helps investors and traders make decisions, but also can be used for auto-trading system. In this article, we have proposed a novel approach of feature selection called "feature importance recap" which combines the feature importance score from tree-based model with the performance of deep learning model. A stacking model is also developed to further improve the performance. Our results shows that proper feature selection approach could significantly improve the model performance, and for financial data, some features have high importance score in many models. The results of stacking model indicate that combining the predictions of some models and feed into a neural network can further improve the performance.

# 1. INTRODUCTION

In modern quantitative trading, data analysis and modelling processes based on statistics, databases, computer science, machine learning, and other quantitative fields have been widely recognized and applied. Effective and accurate prediction for the long and short-term future trends of the financial market is directly related to investors' investment choices and returns.

In the field of data science, studies on deep neural networks have made breakthrough progress in recent years. Deep neural network models such as CNN, RNN, Transformers, etc., are achieving better performance than traditional machine learning in many prediction tasks. However, due to the irregularity, uncertainty, and instability natures of the financial market and the complexity of financial time series data, research on DNN based financial forecasting still has room for improvement.

This project aims to develop an approach to improve the accuracy of future price diction with feature selection and model stacking. Detailed study includes time-series data analysis, feature engineering, DNN model development, feature selection and model stacking.

# 2. RELATED LITERATURE

## 2.1 Background

### 2.1.1 Currency Exchange

Foreign Exchange Market is one of the largest financial market which is active 24/7, and trades continuously from day and night. There are more than 170 currencies in the world, and thousands of currency pairs can be stated in the market. Unlike other financial markets, the foreign exchange market has no specific location and no central transaction. Transactions can be conducted through between banks, enterprises, and individuals. When buying and selling foreign exchange, investors can either earn the exchange rate difference, or convert a foreign currency on hand into another foreign currency with a higher interest rate, thereby obtaining higher interest income. For example, if you buy GBP/USD at 1.5002. When the exchange rate rises to 1.5052, sell the exchange held in your hand and you can make a profit of 50 points.

## 2.1.2 Technical Analysis and Indicators

Technical analysis is a common way to predict price movements. This analysis is based on information stored in the market, which means that it is necessary to refer to the past when predicting future price trends. Traders' forecasts are based on price movements within a certain period. This type of analysis focuses on the composition of charts and formulas to capture major and minor trends, and to identify buying/selling opportunities by estimating the length of the market cycle. The theory of technical analysis is based on three assumptions[1]. Market action discounts everything. 2. Price moves in trends. 3. History repeats itself. Depending on the time span, traders can use intraday (every 5 minutes, every 15 minutes, hourly) technical analysis, weekly or monthly technical analysis[2].

Technical indicators are the elements based on historical trading data such as price, volume, and open interests, rather than the fundamental business indicators like assets, liabilities, or equity to predict the future performance of the market. The basic belief of technical analysis is that history repeats itself and the market movements can be predicted with the help of reams of statistics. Compared to fundamental analysis, technical analysis and indicators focus more on the movements of price, the trading signals, or the changing strength of up and down. Nowadays, technical indicators are widely implemented by investors for short-term investment assistance.

There are four types of indicators applied in our project, momentum, overlap studies, volatility, and price transform.

**Momentum Indicators** measures the movement rate of rise and fall, which means the tendency for rising asset prices to keep increasing, or for the falling ones to keep decreasing. Here are some typical momentum indicators:

- **Moving Average Convergence Divergence (MACD)** [3]MACD is based on the construction principle of moving averages, smoothing the closing price of the gold market price, and calculating the arithmetic average value. Usually, it utilizes two different moving averages and creates the oscillator to illustrate the change from the short time period moving average to the longer time period moving average. The formula for MACD:

$$MACD = \text{shorter time period EMA} - \text{longer time period EMA} \qquad (1)$$

- **Williams %R(WillR)**, shorten of the Williams Percent Range, is a momentum indicator measuring oversold and overbought levels. It is a percentage ranging from 0 to -100, measuring the current close price in comparison to the high-low range over a certain time period. The formula for Williams %R is :

$$\%R = -100 \times \left( \frac{(Highest\ High\ -\ Close\ )}{(Highest\ High\ -\ Lowest\ Low\ )} \right) \qquad (2)$$

When Williams %R runs up to near 0, the close price is very close to the highest high of the certain period, which is the signal of overbought, and that implies the price is in a strong uptrend. Whereas when Williams %R goes down to near -100, the close price nearly reaches the lowest low in the lookback time period. At this moment, the price is in a strong downtrend and has the signal of oversold.

- **BOP** is the shorten of the Balance of Power, which is used to evaluate the struggle between buyers and sellers. The BOP fluctuates around zero, not going up beyond 1 and not going down beneath -1. Positive BOP value represents the buy power is more powerful and negative BOP value implies the strong pressure of selling. When the BOP equals to zero, the power of buyers and sellers neutralizes each other. The formula of BOP is:

$$BOP = SMA\ of\ [\ (Close\ -\ Open\ )/(\ High\ -\ Low\ )] \qquad (3)$$

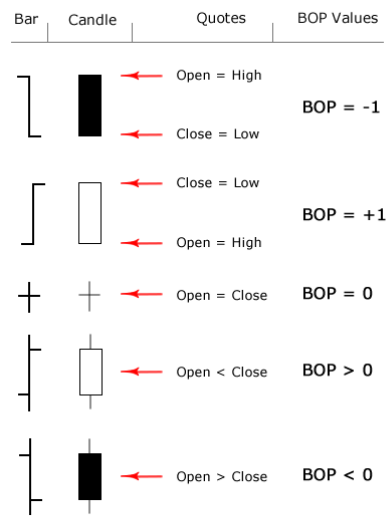

**Figure 1**: *Example of BOP* [4]

**Overlap studies** mostly contain different kinds of moving average indicators. Moving average is one of the most commonly used indicators in technical analysis theory. It is mainly used to confirm, track, and judge the trends and find the buy and sell signals. However, when the market is fluctuating, the moving average will frequently give distorted buy and sell signals.

- **Simple Moving Average (SMA)** means the average mean price in a specific period of time. The formula of SMA is:

$$SMA = [Price\ (1) + Price\ (2) + \cdots + Price\ (n-1) + Price\ (n)]/n \quad (4)$$

- **Exponential Moving Average (EMA)** adds weights to prevent the lagging problem in simple moving average. The formula of EMA is[4]:

$$EMA = EMA\ previous\ + K * (Price - EMA\ previous) \quad (5)$$

Where *K* represents the weight calculated by *K = 2 / (n + 1)*.

- **TRIMA:** Some special moving averages are also included, and we take TRIMA as example. The Triangular Moving Average (TRIMA) is a kind of average price replacing the middle prices of the time period. This method can double-smooth the data using a one-half-length window of the certain time series. The formula for TRIMA is[5]:

$$MA = \frac{SMA(SMAm, Nm)}{Nm}$$

$$N =\ Time\ periods\ + 1$$
$$N_m =\ Round\ (N/2) \quad (6)$$
$$SMAm = (\ Sum\ (\ Price,\ Nm))/Nm$$

**Volatility** indicates the market's expectations and enthusiasm for the stock or the forex. A sharp increase in volatility means that most traders are going to buy or sell during the day, while a decrease in volatility indicates that most traders are not interested in the market. Volatility can be used to measure the volatility of prices to help judge the likelihood of a trend change and the trading atmosphere of the market. Volatility can also be used to help decrease losses and profits.

- **Average True Range (ATR)** [6].Market technicians use the ATR to decide whether enter or exit trades, and ATR is also a useful tool which help the users to understand the volatility of forex of stock by using simple calculations.

Instead of directly indicating the price direction, the ATR can measure volatility caused by gaps and limit up or down moves. The following formula shows how the ATR is calculated[7]:

$$TR = Max\ [(H - L), Abs\ (H - C_P), Abs\ (L - C_P)]$$
$$ATR = \left(\frac{1}{n}\right)\sum_{i=1}^{(n)} TR_i \tag{7}$$

**Price Transform** There are four functions related to the price transform, mainly used to calculate the mean value between the OHLC price:

- **Average Price** The average price is the simplest way to get the average value by calculating the mean value of OHLC price.

$$Average\ Price = (\ Open\ +\ High\ +\ Low\ +\ Close\ )/4 \tag{8}$$

- **Median Price** The Median Price can help users have a quick understand of prices by providing a simple average price on a timestamp. The following formula shows how it is calculated:

$$Median\ Price = (\ High\ +\ Low\ )/2 \tag{9}$$

- **Typical Price** The Typical Price also called the pivot point, is the average price of the stock or forex over the specified period. It is a substitute for the close price, and some technical analysts prefer to use the typical price rather than the close price as the trading price. The following formula shows how it is calculated:

$$Typical\ Price = (\ High\ +\ Low\ +\ Close\ )/3 \tag{10}$$

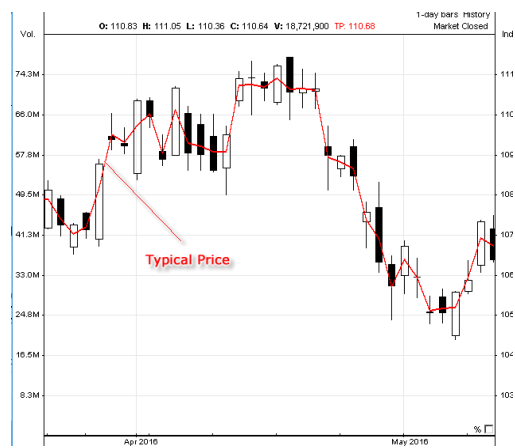

**Figure 2**: *Example of Typical Price [4]*

- **Weighted Close Price** Similar to the median price and the typical price, the weighted closed price is an indicator that is a mean value of each timestamp's price. The reason why it called "weighted close" is that it gives an extra weight on the close price. The following formula shows how to calculate it:

$$Weighted\ Close\ Price\ = \frac{(Close*2)+High+Low}{4} \quad (11)$$

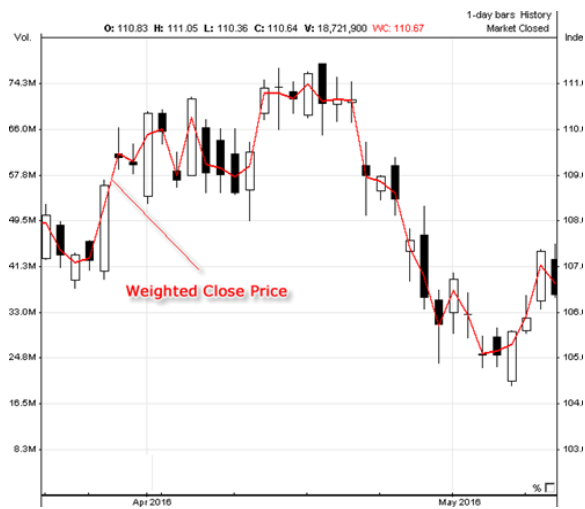

**Figure 3**: *Example of Weighted Close Price [4]*

### 2.1.3 Feature Selection

Feature selection, also known as variable selection or attribute selection. It is a way to select subset of existing attributes for modeling. The main purpose of feature selection is to simplify the model, shorten the training time, avoid the curse of dimensionality, and enhance the generalization capability of the model.

In the feature selection process, if the learner (base learner) is a tree model, the valid features can be selected according to the importance of the features. No matter Random Forest, Gradient Boosting Decision Tree (GBDT), extreme Gradient Boosting (XGBoost) or Light Gradient Boosting Machine (LGBM), they are all based on decision tree models. There are two main ways to calculate the importance of features[8]:

Calculate the importance of each feature during training. During the training process, the importance of features was quantified by recording the total number of splits and the total/average information gain of features. For example, in the actual project, we will use the number of times used in the whole GBDT and XGBoost or the total/average information gain to score the feature importance, and then sort it finally.

Since the Ensemble model itself has a certain randomness when choosing features for splitting, it usually runs multiple models and then ranks them after averaging the importance of features.

Out of Bag (OOB) data was used for calculation after training. To be specific, the OOB data is scored using the trained model to calculate the AUC or other business definition evaluation indicators. Then for each feature in the OOB data:(a) The value of the current feature of random shuffle;(b) Rank the current data and calculate the evaluation indicators;(c) Calculate the change rate of indicators. Following the above approach, a rate of change is obtained for each feature, and the importance of the feature is quantified by ranking the rate of change.

The calculation of feature importance of the tree-based model mentioned above has a common shortcoming, that is, they are sensitive to collinear features. Shapley Additive explanations(SHAP) was designed to mostly mitigate multicollinearity's impact with a very robust statistical foundation[9].The main idea of SHAP value is Shapley value. Shapley value is a method from Coalitional Game Theory. If Shapley value is used to explain the prediction of machine learning, where "total expenditure" is the predicted value of the model for a single instance of the data set. The "player" is the eigenvalue of the instance, and the "revenue" is the actual forecast for that instance minus the average forecast for all instances. The deep learning model itself is a black box in calculation.

## 2.2  Literature Review

Nowadays, the machine learning model is widely used in the quantitative finance industry. However, the historical financial data is noisy and may lead the machine learning models to overfit and impact its performance. Suitable feature engineering and effective forecast combinations can mitigate this risk[10]. Technical indicators are commonly used as input features of the models. Shynkevich and his colleagues elaborated on the feasibility of using technical indicators as feature engineering. They concluded that the input window length and forecast horizon would certainly impact financial forecasting performance[11].

Traditional regression models cannot handle the non-linearity of financial data. Studies have shown that deep neural network methods outperform traditional regression models (OLS, ridgeCV and Bayesian linear regression) in the financial market[12]. Chen et al. have developed an algorithm called Filterbank CNN, which replaces the random-generated filters with the knowledge-based filters. It has been proved that Filterbank could improve the accuracy by up to 15% compared to the traditional rule-based strategy[13].

Since [14] proposed the time series Autoregressive Integrated Moving Average Models(ARIMA) model, the model has been widely used. Especially in financial data forecasting, ARIMA has achieved great results. Based on the basic ARIMA, proposed a fuzzy regression named fuzzy ARIMA. By predicting the historical data of the exchange rate of the NTD and the USD, they proved that this method can not only make better predictions, but also provide decision makers with the best and worst possible scenarios.

Experimental results in [15] demonstrate that ANN based model can closely forecast the forex market. In this paper, they Established and studied six different currency pairs of Australian dollar exchange rate prediction models, namely standard backpropagation (SBP), scaled conjugate gradient (SCG) and backpropagation with Baysian regularization (BPR),and five moving average technical indicators are used to build the models. The results show that all models based on neural networks are better than ARIMA models.

[16] utilize technical indicators on LSTM to make prediction of one-day, three-days and five-days ahead. The experiment results shows that the prediction could generate profit for investors with proper trading strategy.

[17] develops new event-driven features through technical analysis, , which indicate changes in trends in direction. This method largely solves the non-stationarity and high volatility of the foreign exchange market, makes the model interpretable, and the final experiment proves its reliability. With this event-driven method, they implement LSTM, Bi-LSTM and GRU on predication and compared with a basic RNN. They finally got that their best model on 15-minutes interval data for the EUR/GBP currency achieved RMSE 1.5E-3 with a MAPE 0.12% outperforming previous studies.

[18] trained four machine learning ensemble methods: (a) a neural network regression ensemble;(b) a support vector regression ensemble;(c) a boosted regression tree; and (d) a random forest regression. In addition, they established a knowledge base

that based on historical stock data and sentiment analysis of previous news. 19 stocks with different trends are used as a data set. They predict the price of the stock 1-day ahead and eventually get the MAPE ≦ 0.75%.

Due to the interference of noise in financial data, the performance of prediction using machine learning and deep learning models is not ideal[19]. To reduce the impact of the noisy nature of the market, Zezheng Zhang and Matloob Khushi designed the new trading rule features based on technical indicators and trading rules. Also, they proposed a loss function called Sharpe and Sterling Ratio (SSR) which can reduce the variance and the magnitude of drawdowns.

Zhiwen Zeng and Matloob Khushi found that wavelet denoising can also reduce the influence of noise in the financial data and stabilize the data structure[20]. Besides, they used an ARNN (Attention-based Recurrent Neural Network) model integrated ARMIA model to make the prediction and achieved an excellent model performance.

When talking about the wavelet denoising, Yiqi Zhao and Matloob Khushi proposed a Wavelet Denoised-ResNet[21]. And They also combine the model with the LightGBM to get an outstanding model performance in forex price prediction.

Terry and Matloob reviewed all recent stock or forex studies using reinforcement learning as the first deep learning method since 2015[22]. They found that most researches had unrealistic assumptions and showed no meaningful level of profitability. They got the conclusion that the reinforcement learning cannot work as a mature technique utilizing into the real stock and forex market.

Tae and Matloob proposed a novel reinforcement learning model called Deterministic Policy Gradient with 2D Relative-attentional Gated Transformer (DPGRGT), which is a model using creative layers and new reorder methods to stable the reinforcement learning performance[23]. They did back testing on the U.S. stock market of the past 20 years and proved the effectiveness of this model.

[24] demonstrates the feasibility of using technical indicators as features in machine learning to quantify transactions. In their experiment, they took the trading data of 505 S&P 500 stocks in the United States in the past 20 years as historical data and determined 5 technical indicators and 23 basic indicators as input characteristics through analysis. Tests were done on 6 different models (RF, SVM,DT,NB,CNN,KNN) to predict price changes over the next 10 days. As it turned

out, they had an overall accuracy of 83.62%, particularly with their sell signal accuracy of 100%.

In the real financial market, the price of stock has a great relationship with people's production activities. For almost any financial product, news can have a huge impact on the share price. In recent years, because of the great breakthrough of natural language processing, natural language processing has been more and more widely used in quantitative transactions. [25] shows that the text classification in the financial field based on StockTwits text data marked by stock price changes can effectively analyse historical data and easily predict stock trends.

The experiment of [26] proves the reliability of deep learning in financial prediction. They did a lot of paper comparison and analysis, and finally found that the methods combining LSTM with other models can achieve good results.

## 3. PROJECT PROBLEMS

### 3.1 Project Aims & Objectives

This project aims to build a novel system of foreign exchange forecasting based on machine learning and deep learning technologies. Specifically, the. objective is to predict future highest high prices within 5 timestamps of AUD-USD and EUR-USD currency pairs. Precise predictions of the future highest high price are expected to support investment decisions for the greatest financial benefits. As for the innovation of this forecasting system, we mainly focus on two aspects.

First, this project develops a "feature importance recap" approach. As important financial forecasting tools, 71 technical analysis indicators are introduced in this project as the input features of the forecasting models. However, not all indicators are useful for the forecasting target, i.e., Highest price. Therefore, in this project, we develop a feature selection approach to identify a set of most important indicators. This method is combining the "feature importance" approaches of tree-based machine learning models (i.e., XGBoost, Random Forest, and LightGBM) with deep learning experiments (i.e., LSTM and GRU). For each of the 71 technical analysis indicators, a "feature importance coefficient" is assigned.

Second, this project develops a "model stacking" approach. From the above literature reviews, we can find that the existing machine learning and deep learning technologies have made great progress in terms of minimising the losses (indicated by RMSE, MAE, MAPE, etc.) in forex price predictions. To further improve forecasting performance, we look forward to stacking the basic models to build a new complex system, allowing which to learn the advantages from different models, to surpass each single model in terms of prediction performance. Without a doubt, the input features of the stacked models are selected based on the "feature importance recap" approach.

## 3.2 Project Questions

In order to achieve the above project objectives, the project needs to provide explanations for the following questions.

- What is the data source and how to confirm its reliability?

- What data engineering work is required, in terms of the ETL process? Specifically, how to handle missing values, how to standardise data, how to organise data into correct formats for each forecasting model?

- What is the candidate technical analysis indicators (features of model inputs)? How to calculate them?

- As for the tree-based models, in what criterion are the "feature importance" determined?

- What is the threshold to determine whether a feature is important or not?

- What is the candidate forecasting models to build the stacked model?

- What optimisation approaches should be used in each model? Specifically, how to do parameter tuning, how to save model training time?

- What evaluation metrics should be used to measure the success of the prediction?

## 3.3 Project Scope

The project uses M15 history data of AUD-USD pair and EUR-USD pair to predict the future highest high prices within 5 timestamps. We can determine the completion of project by finalising the following tasks.

- In the "feature importance recap" approach: Fitting XGBoost, Random Forest, and LightGBM models for prediction and calculating feature importance;

Based on the obtained feature importance, training LSTM and GRU models and evaluating on out-of-sample data; Determining the recapped feature importance and identifying the final important features; Training the final LSTM and GRU models using the final important features and making evaluation.

- In the "model stacking" approach: Selecting XGBoost, Random Forest, LightGBM, LSTM, and GRU as candidate models; For each subset of the candidate models, applying model stacking approach and making evaluation; Determine the most successful subset of models as the final elements of stacking.

In this project, the prediction loss is measured by root mean square error (RMSE). Other evaluation metrics, i.e., mean absolute error (MAE), and mean absolute percentage error (MAPE) are also considered.

## 4. METHODOLOGIES

### 4.1 Theory and Methods

#### 4.1.1 Long Short Term Memory

LSTM, shorten of Long Short Term Memory, is a special RNN, mainly to solve the problem of gradient disappearance and gradient explosion during long sequence training. Compared to simple ordinary RNNS, LSTM can perform better in longer sequences. LSTM is designed specifically to avoid long-term dependency problems. The default behaviour of the LSTM is to remember the information for a long time without the need for an extra cost.

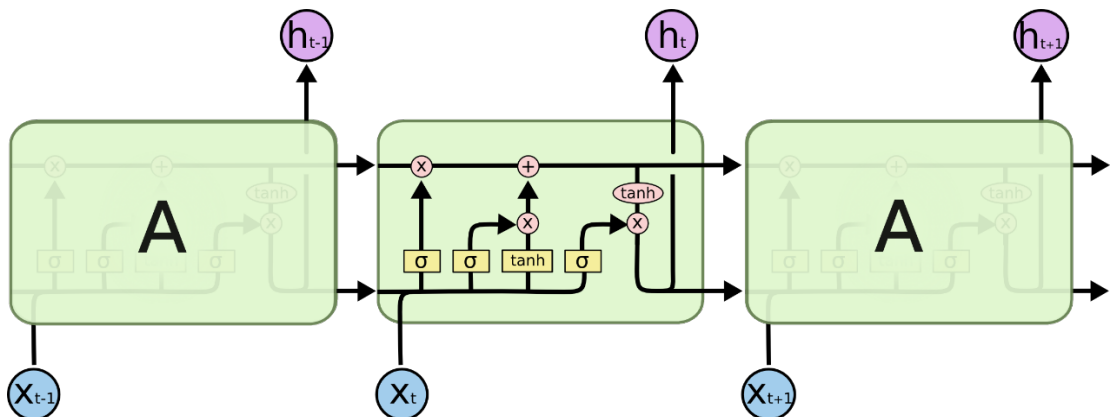

**Figure 4**: *The structure of LSTM* [27]

### 4.1.2 Gate Recurrent Unit

GRU (Gate Recurrent Unit) is a kind of Recurrent Neural Network (RNN). Compared to the LSTM network, the GRU has a simpler structure, and it is easier to train. Therefore, GRU is also a popular RNN model. Similar to LSTM, GRU can also solve the problems of gradient vanish and gradient explosion in the original RNN model.

In LSTM, three gate functions are used: input gate, forgetting gate and output gate. The GRU model is simpler with only two gates: the update gate and the reset gate.

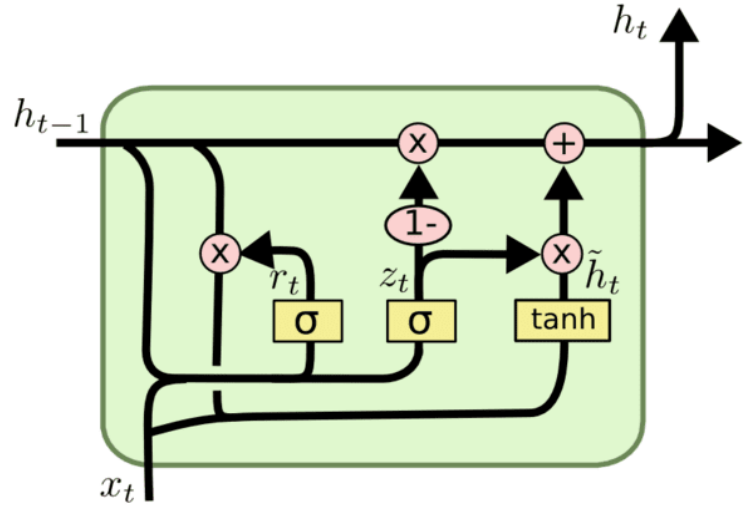

**Figure 5**: *The structure of GRU unit[28]*

The GRU structure is shown in Figure 5. $z_t$ and $r_t$ represent update and reset gates, respectively. The update gate is used to control how much the state information of the previous moment is brought into the current state. The larger the value of the update gate, the more the state information of the previous moment is brought in. The reset gate controls how much information from the previous state is written to the current candidate set; the smaller the reset gate, the less information from the previous state is written. Governing equations of a GRU[29]:

$$\begin{aligned} z &= \sigma(W_z \cdot x_t + U_z \cdot h_{(t-1)} + b_z) \\ r &= \sigma(W_r \cdot x_t + U_r \cdot h_{(t-1)} + b_r) \\ \tilde{h} &= \tanh(W_h \cdot x_t + r * U_h \cdot h_{(t-1)} + b_z) \\ h &= z * h_{(t-1)} + (1 - z) * \tilde{h} \end{aligned} \quad (12)$$

The current input is $x_t$, and the hidden state delivered by the previous node is $h_{t-1}$, which contains information about the previous node.

By combining $h_{t-1}$ with $x_t$, the GRU unit can obtain the output of the current hidden node $y_t$, and the hidden state $h_t$ which will be delivered to the next node.

**4.1.3 XGBoost**

XGBoost is the shortened form of "Extreme Gradient Boosting", which was a scalable tree boosting algorithm published by Tianqi Chen in 2016 [30]. XGBoost greatly improves the speed from the traditional gradient tree boosting system – up to ten times faster than the gbm package. The basic model of XGBoost is decision tree ensembles, which means several trees are ensembled to predict the results.

In decision tree ensembles, the objective function consists of two parts, training loss and regularization term. The formula of objective function can be written as[31]:

$$\text{obj} = \sum_{i=1}^{n} l\left(y_i, \hat{y}_i^{(t)}\right) + \sum_{i=1}^{t} \Omega(f_i) \qquad (13)$$

In this formula, $l$ is the loss function, and $\Omega$ is the regularization term.

The loss function is used to measure how well our predicted value fits the actual training data, while the regularization term represents the complexity of t trees.

For boosting, one tree is added at a time. The new tree is to classify the loss gotten from the last step. If the prediction value at step t is written as $\hat{y}_i^{(t)}$, the formulas for each step are:

$$\begin{aligned}
\hat{y}_i^{(0)} &= 0 \\
\hat{y}_i^{(1)} &= f_1(x_i) = \hat{y}_i^{(0)} + f_1(x_i) \\
\hat{y}_i^{(2)} &= f_1(x_i) + f_2(x_i) = \hat{y}_i^{(1)} + f_2(x_i) \\
&\dots \\
\hat{y}_i^{(t)} &= \sum_{k=1}^{t} f_k(x_i) = \hat{y}_i^{(t-1)} + f_t(x_i)
\end{aligned} \qquad (14)$$

XGBoost uses the greedy algorithm to enumerate all the different tree structures. The approach is to constantly enumerate the structure of different trees, then use the scoring function to find a tree with the best structure, add it to the model, and repeat the operation continuously. The advantage of XGBoost is parallel learning, in which parameters for all leaves are independent and calculated simultaneously. This greatly speeds up the calculation.

Main theories of XGBoost:

- Continuously splitting by features and adding new trees. For every step, a new tree is added to fit the residual of the last prediction.
- When we get k trees after training, we are going to predict the score of a sample. According to the character of this sample, every tree will fall to a corresponding leaf, and each leaf corresponds to a score.
- Finally adding all the correspondent scores of each tree to get the predicted value of the sample.

For every step, the algorithm chooses a new function to minimize the objective value.

$$\text{obj}^{(t)} = \sum_{i=1}^{n} l\left(y_i, \hat{y}_i^{(t)}\right) + \sum_{i=1}^{t} \Omega(f_i) \\ = \sum_{i=1}^{n} l\left(y_i, \hat{y}_i^{(t-1)} + f_t(x_i)\right) + \Omega(f_t) + \text{constant} \quad (15)$$

The novelty of XGBoost is that it uses the second-order Taylor expansion of the loss function to replace the objective function. Here we use Taylor's expansion formula to approximate the objection function:

$$obj^{(t)} = \sum_{i=1}^{n} \left[g_i f_t(x_i) + \frac{1}{2} h_i f_t^2(x_i)\right] + \Omega(f_t) \quad (16)$$

Where $g_i$ represents the first order partial derivative of $\hat{y}_i$, and $h_i$ represents the second order partial derivative of $\hat{y}_i$.. The formula of $g_i$ and $h_i$:

$$g_i = \partial_{\hat{y}_i^{(t-1)}} l\left(y_i, \hat{y}_i^{(t-1)}\right) \\ h_i = \partial^2_{\hat{y}_i^{(t-1)}} l\left(y_i, \hat{y}_i^{(t-1)}\right) \quad (17)$$

In XGboost, the tree function is defined as:

$$f_t(x) = w_{q(x)}, w \in R^T, q: R^d \to \{1, 2, \cdots, T\} \quad (18)$$

Where $w$ represents the vector scores, $q$ is the function to assign points to leaves, and $T$ is the number of leaves. And the complexity function is defined as:

$$\Omega(f) = \gamma T + \frac{1}{2}\lambda \sum_{j=1}^{T} w_j^2 \qquad (19)$$

Let $G_j$ represents the sum of $g_i$ and $H_j$ represents the sum of $h_i$. The objection function can re-formula to this new form:

$$\text{obj}^{(t)} = \sum_{j=1}^{T} \left[ G_j w_j + \frac{1}{2}(H_j + \lambda) w_j^2 \right] + \gamma T \qquad (20)$$

Since this is a quadratic function of $w_j$, the best $w_j$ and the lowest objective reduction can be gotten:

$$w_j^* = -\frac{G_j}{H_j + \lambda}$$
$$\text{obj}^* = -\frac{1}{2} \sum_{j=1}^{T} \frac{G_j^2}{H_j + \lambda} + \gamma T \qquad (21)$$

Where $G_j, H_j, \lambda$, and $\gamma$ are the parameters for the models to learn. Compared to the traditional random forest method, the learning parameters squeeze a lot.

### 4.1.4 Random Forest

Random forest is a method of Bagging in ensemble learning. It is composed of a large number of decision trees.

- The decision tree:

    Each decision tree in the random forest is independent, they will use several features in the dataset as the input of the decision tree. Commonly used decision trees include ID4, C4.5, CART and so on. In the process of spanning tree, which feature needs to be selected for subdivision. The principle of selection is to improve the purity as much as possible after separation, which can be measured by indicators such as information gain, gain rate and Gini coefficient. If it's a tree, pruning should be performed to avoid overfitting, removing nodes that might cause overfitting in the validation set.

- The principle of random forest:

    When the random forest algorithm is used for classification or regression, each decision tree in the forest will be calculated independently and a

prediction result will be obtained. After that, the algorithm will collect and count the prediction results of all the decision trees and select the prediction results that appear the most times in the decision tree as the final result.

- Advantages and disadvantages:

Random forest has many advantages, it can be used on the data with numerous features. When using the random forest algorithm，it is unnecessary to reduce the dimensionality of the data or do feature selection. In addition, the random forest can also obtain the feature importance, which is the reason that we use it in our feature importance recap part. In addition, it is not easy to overfit, and the computational cost is not acceptable. Random forests also have some drawbacks. It is easy to overfit in the dataset containing noise.

- Application:

Random forest can be used in regression and classification problems, as well as the clustering problem of unsupervised learning and outlier detection.

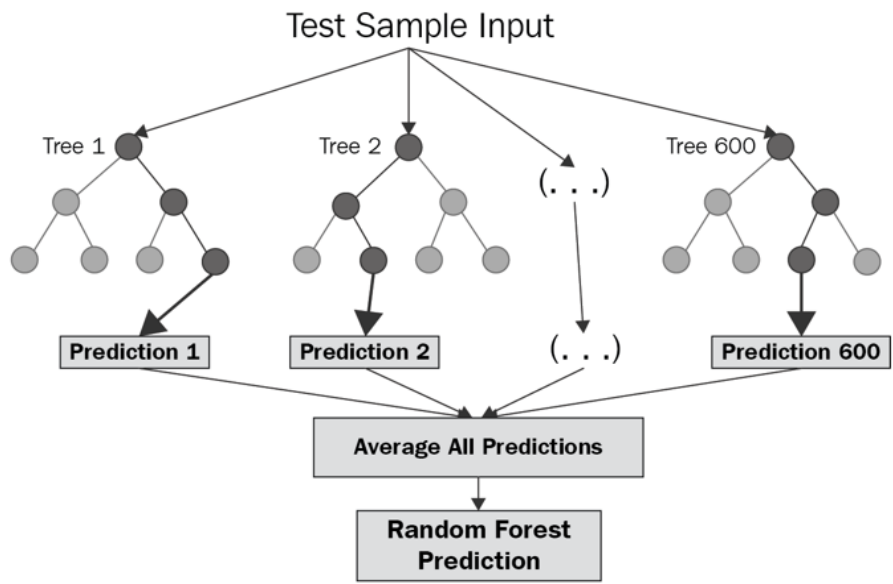

**Figure 6**: *An example of random forest [32]*

### 4.1.5 LightGBM

LightGBM, shorten of Light Gradient Boosting Machine, is a gradient boosting framework focusing on speeding up the training. Compared to XGBoost, Light GBM grows leaf-wise trees and uses a technique called Gradient-based One-Side Sampling (GOSS) to filter out the split data. As a result, Light GBM can run almost seven times faster than XGBoost and can deal with larger datasets. Using GOSS can reduce a large number of data instances with only small gradients, so that when calculating information gain, only the remaining data with high gradients can be used. Compared to XGBoost traversing all feature values, it saves a lot of time and space.

LightGBM has the following features：

- Histogram algorithm:

    Compared to the pre-sorted algorithm, the memory occupation and the data separation complexity of histogram algorithm is lower. The idea is to discretise continuous floating-point features into k discrete values, and construct a histogram with a width of k. Then the algorithm traverses the training data and counts the cumulative statistics of each discrete value in the histogram. The feature selection step is only to find the optimal segmentation point according to the largest value of the histogram.

- Leaf-wise growing strategy:

    LightGBM adopts a leaf-wise growing strategy, which finds the leaf with the largest split gain (generally the largest amount of data) from all the current leaves. Therefore, compared with level-wise strategy, the leaf-wise can reduce more errors and get better accuracy when the number of splits is the same. The disadvantage of leaf-wise is that it may grow a deeper decision tree, resulting in overfitting. To avoid this, LightGBM adds a maximum depth limit on the top of leaf-wise so as to prevent overfitting and ensure high efficiency.

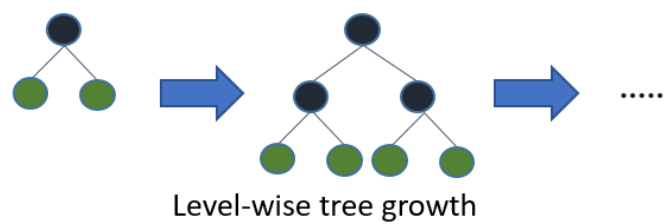

Level-wise tree growth

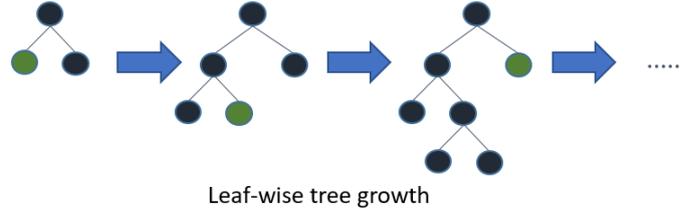

Leaf-wise tree growth

**Figure 7**: *An example of LightGBM [33]*

- Feature Parallelization:

    LightGBM supports parallel learning, including feature parallelization and data parallelization. The main idea of feature parallelization is to find the optimal segmentation points on different feature sets, and then synchronize the optimal segmentation points between the machines. Data parallelization allows different machines to construct histograms locally, then merge them globally, and finally find the optimal split point on the merged histogram.

### 4.1.6 ARIMA

ARIMA (autoregressive integrated moving average) is a famous time series method proposed by box and Jenkins in 1970s. It is a complex statistical model which is the auto regression model integrated with the moving average model. And P, D, Q are the three essential input parameters of this model. P is the number of lag orders in the AR model. D is the degree of differencing. Q is the size of the moving average window.

ARIMA is a generalization of the ARMA(autoregressive moving average) model. The ARMA*(p, q)* model is given by[34]:

$$\left(1 - \sum_{i=1}^{p'} \alpha_i L^i\right) X_t = \left(1 + \sum_{i=1}^{q} \theta_i L^i\right) \varepsilon_t \qquad (22)$$

where $a_i$ are the parameters of the AR model, the $\theta_i$ are the parameters of MA model, $L$ is the operator of lag, and the $\varepsilon_t$ are the error terms.

Arima can be used when the time series data are stable. Stationary time series represent time series without trend, one of which has constant mean value and variance over time.

The autocorrelation function (ACF) and partial correlation function (PACF) are used to determine the values of P and Q in autoregression (AR) and moving average (MA), or we can use the criterion called Akaike information criterion (AIC) to make sure the input parameters. In this project, we use AIC to calculate the appropriate input parameters.

AIC is written as:

$$AIC = -2\log(L) + 2(p + q + k) \qquad (23)$$

where p is the order of the AR part and q is the order of the MA part, and the k is the intercept and $L$ is the likelihood.

## 4.2 Data Processing

### 4.2.1 Data Collection

The forex data used in our project was extracted from Oanda.com, which is one of the biggest forex trading brokers in the world. Moreover, the history trading prices data was downloaded with Python code by calling the API provided by Oanda. Two kinds of currency transactions, including AUD-USD and EUR-USD, are chosen as our datasets, and the data fields of the datasets are timeframe and OHLC prices. These two datasets range from June 2014 to April 2021.

### 4.2.2 Target Generation

Instead of predicting the close price of the next timestamp, this project aims to predict the maximum value of the high prices on the next five timestamps, which can be called the "highest-high" price in short. The reason why we predict the "highest-high" price, rather than the close price, is that the prediction of the "highest-high" is more valuable and meaningful, which can help investors estimate the upper limit of the future growth and make a more appropriate trading strategy. The formula for calculating the "highest-high" price is as below:

$$HighestHigh|_{T_i} = max(High|_{T_i+1}, High|_{T_i+2}, High|_{T_i+3}, High|_{T_i+4}, High|_{T_i+5}) \qquad (24)$$

We calculate the corresponding "highest-high" value for all timestamps of the dataset, and the "highest-high" price will be used as the label of the dataset to train the model and get the model performance.

### 4.2.3 Feature Engineering

For feature engineering, we first generate technical indicators for all timestamps of the dataset by using the TA-Lib (Technical Analysis Library) package, which is a free open-source library. The reason why we generate technical indicators is that some of them can reveal the pattern of the historical data and have a positive effect on the prediction, so as to improve the performance of models.

We calculate 139 technical indicators from five types of functions of the TA-Lib, including "overlap studies", "momentum", "volatility", "cycle" and "pattern recognition" for all timestamps of the dataset. However, some indicators are meaningless and invalid for this dataset, and their values are always zero. Therefore, we discard invalid indicators, and 71 indicators are left shown in Figure 8 as the input features of models.

| Momentum | ADX<br>- Average Directional Movement Index | Overlap studies | BBANDS<br>- Bollinger Bands |
|---|---|---|---|
| | ADXR<br>- Average Directional Movement Index Rating | | DEMA<br>- Double Exponential Moving Average |
| | APO<br>- Absolute Price Oscillator | | EMA<br>- Exponential Moving Average |
| | AROON<br>- Aroon | | HT_TRENDLINE<br>- Hilbert Transform - Instantaneous Trendline |
| | AROONOSC<br>- Aroon Oscillator | | KAMA<br>- Kaufman Adaptive Moving Average |
| | BOP<br>- Balance Of Power | | MA<br>- Moving average |
| | CCI<br>- Commodity Channel Index | | MAMA<br>- MESA Adaptive Moving Average |
| | CMO<br>- Chande Momentum Oscillator | | MAVP<br>- Moving average with variable period |
| | DX<br>- Directional Movement Index | | MIDPOINT<br>- MidPoint over period |
| | MACD<br>- Moving Average Convergence/Divergence | | MIDPRICE<br>- Midpoint Price over period |
| | MACDEXT<br>- MACD with controllable MA type | | SAR<br>- Parabolic SAR |
| | MACDFIX<br>- Moving Average Convergence/Divergence Fix 12/26 | | SAREXT<br>- Parabolic SAR - Extended |
| | MINUS_DI<br>- Minus Directional Indicator | | SMA<br>- Simple Moving Average |
| | MINUS_DM<br>- Minus Directional Movement | | T3<br>- Triple Exponential Moving Average (T3) |
| | MOM<br>- Momentum | | TEMA<br>- Triple Exponential Moving Average |
| | PLUS_DI<br>- Plus Directional Indicator | | TRIMA<br>- Triangular Moving Average |
| | PLUS_DM<br>- Plus Directional Movement | | WMA<br>- Weighted Moving Average |
| | PPO<br>- Percentage Price Oscillator | Volatility | ATR<br>- Average True Range |
| | ROC<br>- Rate of change : ((price/prevPrice)-1)*100 | | NATR<br>- Normalized Average True Range |
| | ROCP<br>- Rate of change Percentage: (price-prevPrice)/prevPrice | | TRANGE<br>- True Range |
| | ROCR<br>- Rate of change ratio: (price/prevPrice) | Price transform | AVGPRICE<br>- Average Price |
| | ROCR100<br>- Rate of change ratio 100 scale: (price/prevPrice)*100 | | MEDPRICE<br>- Median Price |
| | RSI<br>- Relative Strength Index | | TYPPRICE<br>- Typical Price |
| | STOCH<br>- Stochastic | | WCLPRICE<br>- Weighted Close Price |
| | STOCHF<br>- Stochastic Fast | NA | |
| | STOCHRSI<br>- Stochastic Relative Strength Index | | |
| | TRIX<br>- 1-day Rate-Of-Change (ROC) of a Triple Smooth EMA | | |
| | ULTOSC<br>- Ultimate Oscillator | | |
| | WILLR<br>- Williams' %R | | |

**Figure 8**: *A technical indicators used as features*

ARMIA(Autoregressive integrated moving average) is a statistical model which is widely used in time series data prediction.

In our project, as shown in Figure 9, although ARIMA prediction results are very close to the actual value and the RMSE value of this model is tiny. It has the problem of lagging, which means all the prediction results of the predicted timestamps just like a copy and paste to the value of the previous timestamp. Therefore, we choose the prediction of the ARIMA model as an input feature of the model to help improve the model performance.

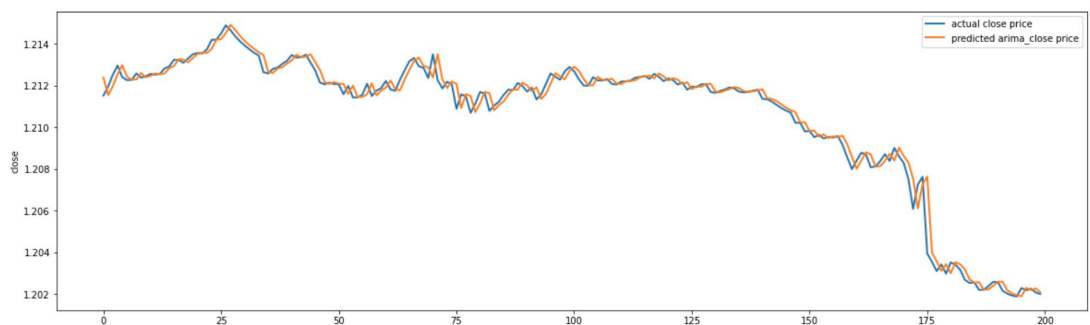

**Figure 9**: *The visualization of the ARIMA prediction result.*

### 4.2.4 Data Transformation

Deep learning and machine learning models can be used for forex prediction, which is a time-series prediction problem. However, since these models only accept datasets from supervised learning, the time-series problem needs to be transformed into a supervised learning one. In our project, two methods are respectively used to carry out data transformation:

- In order to be accepted by the deep learning models, including the LSTM and GRU, the dataset will be directly converted into the type of supervised learning by using one Keras API under pre-processing called: timeseries_dataset_from_array. And the previous five timestamps are used to predict the "highest-high" price of the next timestamp.

- For machine learning models, including XGBoost, Random Forest, and LightGBM, we write a function by ourselves to transfer the data. And similar to the data transformation in deep learning，the previous five timestamps are used to predict the highest-high price of the next timestamp.

| Index | datetime | Input features | | | | | | | | | Label |
|---|---|---|---|---|---|---|---|---|---|---|---|
| | | open(t-4) | high(t-4) | low(t-4) | close(t-4) | upperband(t-4) | ... | medprice(t) | typprice(t) | wclprice(t) | highest_high |
| 4 | 2014/4/1 21:45 | 0.92424 | 0.92442 | 0.9236 | 0.92386 | 0.924807 | ... | 0.92432 | 0.92428 | 0.92426 | 0.92461 |
| 5 | 2014/4/1 22:00 | 0.92387 | 0.92426 | 0.9239 | 0.92402 | 0.924375 | ... | 0.924355 | 0.924413 | 0.924442 | 0.92501 |
| 6 | 2014/4/1 22:15 | 0.92402 | 0.92405 | 0.9236 | 0.92362 | 0.924445 | ... | 0.92423 | 0.924147 | 0.924105 | 0.92501 |
| 7 | 2014/4/1 22:30 | 0.92364 | 0.92447 | 0.9236 | 0.92446 | 0.924624 | ... | 0.923755 | 0.92378 | 0.923792 | 0.92501 |
| 8 | 2014/4/1 22:45 | 0.92445 | 0.92445 | 0.9242 | 0.9242 | 0.924605 | ... | 0.92368 | 0.923707 | 0.92372 | 0.92501 |
| ... | ... | ... | ... | ... | ... | ... | ... | ... | ... | ... | ... |
| 124046 | 2019/3/31 21:45 | 0.71132 | 0.71132 | 0.711 | 0.71108 | 0.712408 | ... | 0.71077 | 0.710707 | 0.710675 | 0.71164 |
| 124047 | 2019/3/31 22:00 | 0.71105 | 0.71114 | 0.7108 | 0.71094 | 0.711767 | ... | 0.71065 | 0.7107 | 0.710725 | 0.71164 |
| 124048 | 2019/3/31 22:15 | 0.7109 | 0.71107 | 0.7108 | 0.71106 | 0.711388 | ... | 0.710875 | 0.71091 | 0.710927 | 0.71164 |
| 124049 | 2019/3/31 22:30 | 0.71103 | 0.71117 | 0.711 | 0.71104 | 0.711311 | ... | 0.71124 | 0.71133 | 0.711375 | 0.71194 |
| 124050 | 2019/3/31 22:45 | 0.711 | 0.711 | 0.7105 | 0.71058 | 0.711313 | ... | 0.711405 | 0.71135 | 0.711323 | 0.712 |

**Figure 10**: *The dataset after transformation for the machine learning models*

These two data conversion methods are essentially the same: by using the sliding window method, all columns of the past five timestamps are taken as the input features of the model, and the highest high price of the timestamp is taken as the label of the model. For example, as shown in Figure 10, if we choose the highest-high price of timestamp "t" as the label in the model, the input features are all columns of timestamp "t-4" combined all columns of timestamp "t-3" and all columns of timestamp "t-2" and all columns of timestamp "t-1" and all columns of timestamp "t".

### 4.2.5 Data cleaning

In the prediction of the "highest-high" price and the generation of technical indicators, some invalid rows in the data need to be discarded.

- In the prediction of "highest-high" price, because the last five timestamps cannot obtain the high price of their future five timestamps, it is impossible for them to calculate their respective "highest high" price and their corresponding rows need to be discarded.

- In the generation of technical indicators, some indicators for the number of rows do not exist, and the corresponding rows need to be discarded. For example, in calculating the indicator called the moving average 10(MA10), the MA10 value of the first nine timestamps cannot be calculated, and these rows need to be discarded.

### 4.2.6 Data splitting

The entire dataset, which contains data from June 2014 to April 2021, is used in machine learning and deep learning models and subsequent stacking systems. And the dataset is divided into three parts: the training set, validation set, and test set. The training set, which is used for model training, ranges from June 2014 to May 2019. The validation set used for parameter tuning of models ranges from June 2019 to May 2020. The test set ranges from Jun 2020 to April 2021, and it is used for model evaluation and also used in the subsequent stacking system. In the model evaluation, each model uses the data from the test set to predict and obtain a prediction result with the same row number as the test set. The prediction of each model will be used as the input feature in the subsequent stacking system.

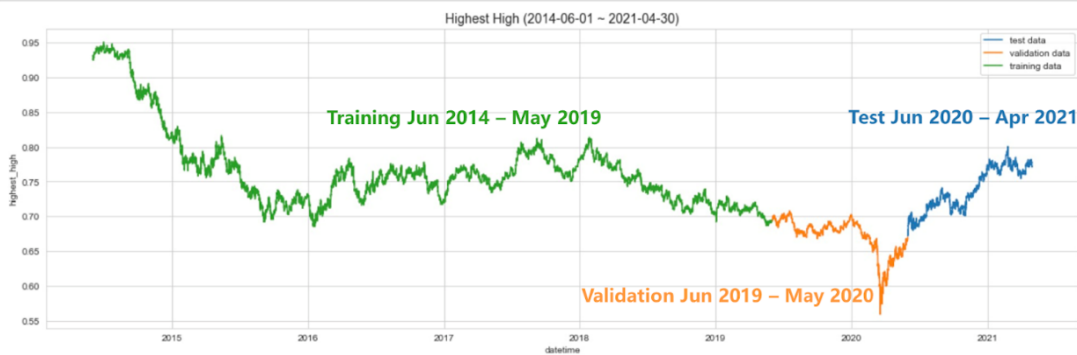

**Figure 11**: *The visualization of data splitting for AUD_USD dataset*

## 4.3    Traditional feature importance methods

The purpose of feature importance is to highlight the correlation between the features and the target, which can help us understand the data set and the model. The importance of a feature can be expressed by its feature importance score. The feature importance score is usually calculated by a prediction model fitted to the data set, providing insight into this model and how important the feature is when making predictions. Based on the feature importance result, we can further improve the model performance by dropping the features with low importance score. Hence, the dataset dimension is reduced which will also speed up the training process.

For a tree-based model, the importance score measures the value of the feature in the tree construction. The more a feature is used to build a tree in the model, the higher its importance is. Feature importance is obtained by calculating and sorting each feature in the data set. In a single decision tree, the feature importance is calculated by the amount of performance improvement for each feature split point, and the weight and frequency of making improvement is based on the node. In other words, the greater the improvement performance of a feature on the split point (the closer it is to the root node), the greater the weight; the more boosted trees are selected, the more important the feature is. The performance measurement can be entropy gain, Gini purity, permutation or SHAP（Shapley Additive explanation）.

In this project, we have applied built in feature importance functions for XGBoost, Random Forest and LightGBM. For XGB, the feature importance is based on the average gain across all splits the feature is used in. Random forest used the Gini importance (mean decrease impurity). In random forest, the features for internal nodes are selected by its variance reduction. The importance score is the average of how a feature decrease the impurity over all trees in the forest. In LightGBM, the score is the count of how many times a feature is used in splits.

### 4.4 Feature Importance Recap

Feature importance result from a model is for the particular model only. For example, the same feature will have different score and rank in different models. Therefore, we cannot simply apply the feature importance to other models without any modification. For a tree-based model, it is easy to calculate its importance using the methods we discussed in previous section. For deep learning model, as most of them are block-box model, we cannot tell which feature is more importance. We could use SHAP （Shapley Additive explanation） to explain the model. But there are several limitations:

1. The calculation is time-consuming, its complexity is increased on an exponential level if we add more features in the dataset. For financial data, we could easily calculate hundreds of technical indicators as features.

2. SHAP result of a feature is not based on the performance of the model after removing it.

3. If there are correlations between features, this perturbation-based method may produce unrealistic data. In financial data, as the technical indicators are calculated with financial equations, there could be features with high correlations but still useful.

Therefore, we proposed a feature selection approach called 'Feature importance recap'. The main theory is that, as the tree-based model is easier to explain, we will apply the result from these models and use the selected features as input for deep learning model. In this project, we tested the feature importance with XGB, random forest and LightGBM and make prediction with LSTM and GRU.

The first step is to split the training dataset into train and test dataset for feature importance calculation. We used data from April 2014 to March 2018 as new training set and April 2020 to April 2021 as testing set. Then we calculate the feature importance score with XGBoost, Random Forest and LightGBM.

In the second step, we use the three feature sets from step one and apply deep learning models to get the three RMSE result. Then we normalize the feature importance score from step one with MinMaxScaler and divide the importance score by the corresponding RMSE. Hence the result will represent both the importance score and its performance in deep learning model. The final importance score is the sum of the importance score from all three tree-based models.

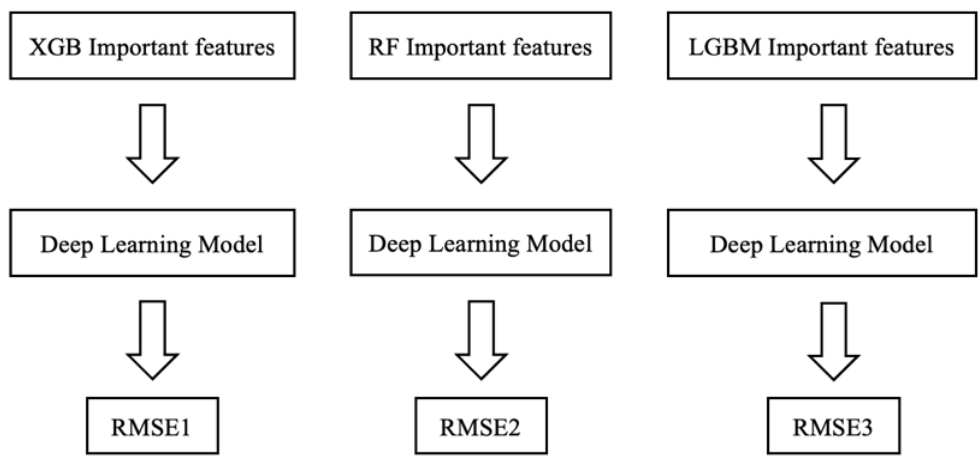

**Figure 12**: *Calculation of final feature importance score*

$$\text{Final feature score} = \text{SUM} \left( \frac{XGB \text{ score}}{RMSE1}, \frac{RF \text{ score}}{RMSE2}, \frac{LGBM \text{ score}}{RMSE3} \right) \quad (25)$$

## 4.5   Model stacking

Stacking is a method from ensemble learning which can be used for both regression and classification problems. Stacked models win most machine learning and data science contests since the stacking system can improve the model performance to a new level by mixing multiple existing models.

As shown in the stacking architecture of Figure 13, the stacking structure has two layers. In the first layer, several different models are used to generate the predictions. Each model's prediction is used as an input feature to form a data frame that will be passed to the second layer of stacking. Finally, in the second layer, a model is used to combine the results from columns of the data frame and get the final prediction, which tends to have a better performance.

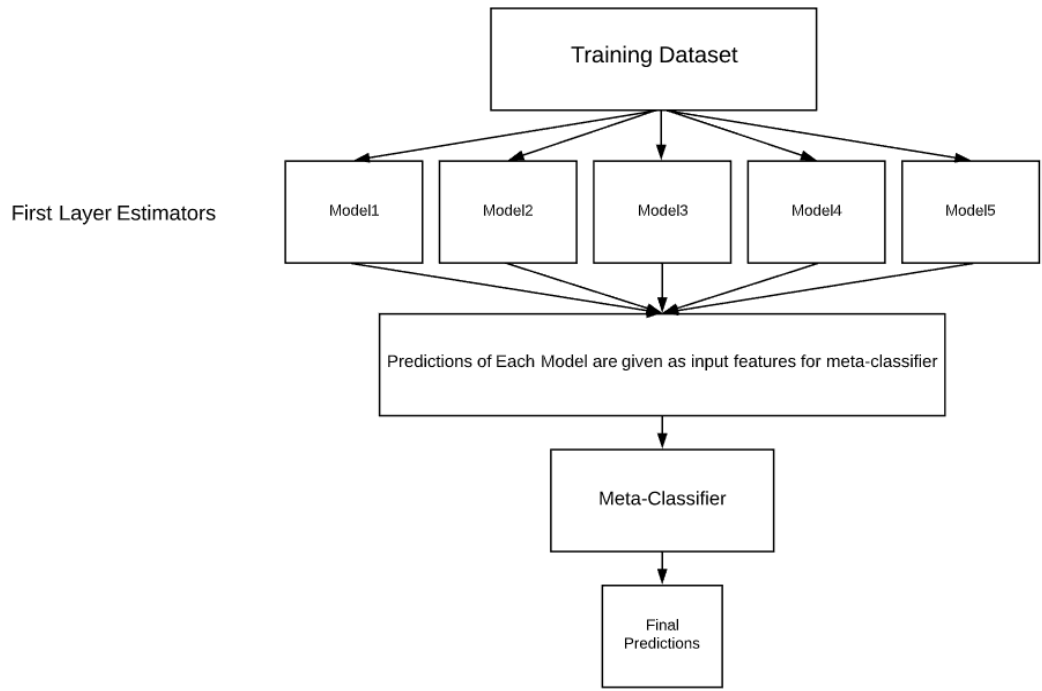

**Figure 13**: *Stacking architecture [35]*

In our project, there are five models in the first layer of the stacking system: three of them are machine learning models, including LightGBM, Random Forest, and XGBoost. Two are deep learning models, including LSTM and GRU. In the first layer of stacking, five models are used respectively to predict the "highest-high" price of the test set(June2020-April2021), and the prediction result is obtained respectively. Each prediction result is used as a column to form a data frame, which has five columns corresponding to the prediction results of the five models. Then put the set of the actual "highest-high" price as the sixth column in the data frame, which is the label of the dataset. Then, this data frame is delivered to the stacking second layer as the dataset.

In the second layer of the stacking system, a neural network model is used to combine the results of the five models and obtain the final prediction. Before inputting the neural network model(NN model), we first carry out data splitting on the dataset delivered by the first layer. The dataset is divided into three parts: the training set, validation set, and test set. The training set used for NN model training ranges from June 2020 to October 2020. The validation which is used for parameter tuning of the NN model ranges from November 2020 to December 2020. The test set ranges from January 2020 to April 2021, which is used to obtain the predicted results and stacking model evaluation.

When it comes to this NN model, it is not fixed and has 31 possible structures. Because using all the predictions from five models as input features may not be the optimal solution. It is possible to use the prediction results of four models as input of the NN model to achieve better performance. Therefore, it is a combination problem that has five candidate models.

As Figure 14 shows, there are 31 possibilities of combinations. We will compare the performance of these 31 models and finally choose a model with the smallest RMSE as the final NN model of the second layer of stacking. And the NN model prediction results are the final prediction results of the stacking system.

| COMBIN(5,1) = 5 | Xgboost | COMBIN(5,3) = 10 | Xgboost, LightGBM , RF |
|---|---|---|---|
| | LightGBM, | | Xgboost, LightGBM , LSTM |
| | RF | | Xgboost, LightGBM,GRU |
| | LSTM | | Xgboost, RF,LSTM |
| | GRU | | Xgboost, RF,GRU |
| | Xgboost, LightGBM, RF, LSTM | | Xgboost,LSTM,GRU |
| | Xgboost, LightGBM, RF, GRU | | LightGBM, RF,LSTM |
| | Xgboost, LightGBM, LSTM, GRU | | LightGBM, RF,GRU |
| | Xgboost, RF, LSTM, GRU | | LightGBM, LSTM, GRU |
| COMBIN(5,4) = 5 | LightGBM, RF, LSTM, GRU | | RF, LSTM,GRU |
| COMBIN(5,5) = 1 | Xgboost, LightGBM, RF, LSTM, GRU | | N/A |

**Figure 14**: *31 combinations from the five models*

## 4.6  Evaluation

Because the primary goal of our study is to establish a model that can accurately predict the future forex price, and it is a regression problem. Evaluation metrics of regression models mainly include MSE, MAE, MAPE, and RMSE.

### 4.6.1 Mean squared error (MSE)

The mean squared error (MSE) indicates the difference of the regression predictions with actual results. By calculating the distances from the actual points to the regression predictions and squaring them. The range of the MSE is $[0, +\infty)$, and the lower MSE always means a better model performance.

$$MSE = \frac{1}{n}\sum_{i=1}^{n}(\hat{y}_i - y_i)^2 \qquad (26)$$

### 4.6.2 Root mean squared error (RMSE)

The RMSE is a measure of the differences between predicted results the values observed. RMSE is non-negative, and the range of it is $[0, +\infty)$. When the RMSE is equal to zero, the prediction results perfectly fit the actual data. and the lower RMSE indicates a better model performance.

$$RMSE = \sqrt{\frac{1}{n}\sum_{i=1}^{n}(\hat{y}_i - y_i)^2} \qquad (27)$$

### 4.6.3 Mean absolute error (MAE)

Mean absolute error (MAE) measures errors between the predicted results and the actual results. The MAE can avoid the problem of error cancellation, so it can accurately reflect the actual prediction error. MAE calculates the average of the residuals directly, while RMSE penalizes more for high differences than MAE.

$$MAE = \frac{1}{n}\sum_{i=1}^{n}|\hat{y}_i - y_i| \qquad (28)$$

### 4.6.4 Mean Absolute Percentage Error (MAPE)

The MAPE, which is also called mean absolute percentage deviation (MAPD), is a measure of prediction accuracy of regression in statistics.

$$MAPE = \frac{100\%}{n}\sum_{i=1}^{n}\left|\frac{\hat{y}_i - y_i}{y_i}\right| \qquad (29)$$

As shown in the MAPE formula, $y_i$ is the actual result of the ith sample, and $\hat{y}_i$ is the result of the ith sample predicted by the model. The range of MAPE is $[0, +\infty)$, and when MAPE is 0, the model is the perfect one. If the MAPE value is more than 1, it means the model performed badly. Although the concept of MAPE is intuitive and convincing, it has a very significant disadvantage: if there is any zero-value contained in the actual value of the dataset, the MAPE cannot be used.

# 5. RESULTS

## 5.1 Feature importance recap result

The testing dataset is from Jun 2020 to April 2021 with timeframe of 15 minutes of AUD_USD and EUR_USD. We use the previous 5 timestamps to predict the highest high price in the next 5 timestamps. The model performance is measured with RMSE, MAE and MAPE. We have tested with top 20 and top 30 features. The best result for AUD_USD is GRU with top 20 features, and for EUR_USD is LSTM with top 30 features. The result with feature importance recap is very close.

The hyperparameters are set as batch_size = 256 for timeseries dataset and learning rate is 1E-5.

| Input Features | RMSE (E-3) | MAE (E-3) | MAPE (E-3) |
|---|---|---|---|
| All features | 2.1687 | 2.5486 | 3.4228 |
| XGB 20 | 1.4016 | 1.0578 | 1.4399 |
| LGBM 20 | 1.8481 | 1.4253 | 1.9312 |
| RF 20 | 1.7717 | 1.337 | 1.8289 |
| Final 20 | 0.8037 | 0.6491 | 0.883 |
| XGB 30 | 1.5245 | 1.169 | 1.5943 |
| LGBM 30 | 1.2218 | 0.8962 | 1.2205 |
| RF 30 | 1.6558 | 1.2519 | 1.7056 |
| Final 30 | 0.9482 | 0.6762 | 0.9195 |

**Figure 15**: *Test result with LSTM on AUD_USD*

| Input Features | RMSE (E-3) | MAE (E-3) | MAPE (E-3) |
|---|---|---|---|
| All features | 3.1101 | 2.6433 | 3.5785 |
| XGB 20 | 1.0547 | 0.7771 | 1.0793 |
| LGBM 20 | 1.3268 | 0.9391 | 1.3382 |
| RF 20 | 1.2951 | 0.9605 | 1.2946 |
| Final 20 | **0.7563** | **0.5312** | **0.7223** |
| XGB 30 | 1.3656 | 1.0436 | 1.4417 |
| LGBM 30 | 1.0056 | 0.7209 | 0.979 |
| RF 30 | 1.1578 | 0.8649 | 1.1754 |
| Final 30 | 0.8412 | 0.5948 | 0.8065 |

**Figure 16**: *Test result with GRU on AUD_USD*

| Input Features | RMSE (E-3) | MAE (E-3) | MAPE (E-3) |
|---|---|---|---|
| All features | 2.03 | 1.8007 | 1.5104 |
| XGB 20 | 0.9664 | 0.637 | 0.539 |
| LGBM 20 | 1.2608 | 1.0072 | 0.8457 |
| RF 20 | 1.1261 | 0.7749 | 0.6553 |
| Final 20 | 0.8693 | 0.6497 | 0.5492 |
| XGB 30 | 1.7824 | 1.4891 | 1.2469 |
| LGBM 30 | 1.9033 | 1.6794 | 1.4081 |
| RF 30 | 2.3868 | 2.1097 | 1.77 |
| Final 30 | **0.7945** | **0.5449** | **0.461** |

**Figure 17**: *Test result with LSTM on EUR_USD*

| Input Features | RMSE (E-3) | MAE (E-3) | MAPE (E-3) |
|---|---|---|---|
| All features | 3.1621 | 2.8576 | 2.3957 |
| XGB 20 | 1.97 | 1.7413 | 1.4597 |
| LGBM 20 | 0.9296 | 0.6409 | 0.5402 |
| RF 20 | 1.5558 | 1.2992 | 1.092 |
| Final 20 | 0.8824 | 0.7141 | 0.6048 |
| XGB 30 | 1.1294 | 0.8719 | 0.7336 |
| LGBM 30 | 1.3816 | 1.1233 | 0.9436 |
| RF 30 | 1.1442 | 0.6409 | 0.5402 |
| Final 30 | 0.9444 | 0.6538 | 0.5514 |

**Figure 18**: *Test result with GRU on EUR_USD*

More experiment results with selected features please refer to the appendix and the artifacts.

The result shows that the experiments with feature selection outperforms the experiments fed with all features. With the feature selected with "feature importance recap" the performance is even better.

| | Selected features |
|---|---|
| XGB 30 23 | ['close','high','typprice','sma10','dema','wclprice','low','tema','upperband','avgprice','arima_high','sma30','open','kama','ht','midprice','midpoint','t3','wma5','lowerband','medprice','trima30','atr'] |
| LGBM 30 27 | ['close','high','trange','wclprice','upperband','atr','dema','trima30','bop','willr','fastk','natr','plus_dm','medprice','adxr','macdsignalfix','midpoint','typprice','macdsignalext','tema','slowd','ht','mom','minus_di','macdhistext','sma30','fastkrsi'] |
| RF 30 19 | ['close','wclprice','high','avgprice','dema','medprice','typprice','low','open','lowerband','kama','midprice','ema','arima_close','wma5','midpoint','tema','ma','arima_high'] |
| Final 30 28 | ['close','high','wclprice','trange','typprice','dema','upperband','atr','trima30','bop','sma10','willr','medprice','natr','fastk','plus_dm','adxr','midpoint','macdsignalfix','macdsignalext','avgprice','tema','slowd','ht','mom','minus_di','sma30','macdhistext'] |

**Figure 19**: *Selected features for LSTM on AUD_USD*

By investigating the selected features, we found that some features like close, high, price transform indicators, are selected in all experiments. Figure 19 is a sample of selected features for LSTM on AUD_USD. As the features are selected with supervised dataset, some features are selected twice. For example, both close(t) and close(t-1) are selected. The timeseries dataset will use all previous values of the feature so the selected feature set will be less than 20 or 30. The overlapping among the selected features will be included in future research.

## 5.2 Model stacking result

On the unseen data set, we test the performance of stacking for two different currency transactions, including AUD-USD and EUR-USD. By comparing the performances of 31 combinations, we choose the most suitable stacking model, which has the smallest RMSE.

The performances of 31 stacking combinations are shown in Figure 20. The stacking system using XGBoost, LightGBM, and GRU has the smallest RMSE overall. The RMSE of this stacking system is 6.799e-04, which is much better than RMSE 7.563e-04 of GRU, the single best one.

Hence, in the dataset of M15 AUD-USD, we finally choose to use XGBoost, LightGBM, and GRU in the first layer of the stacking system.

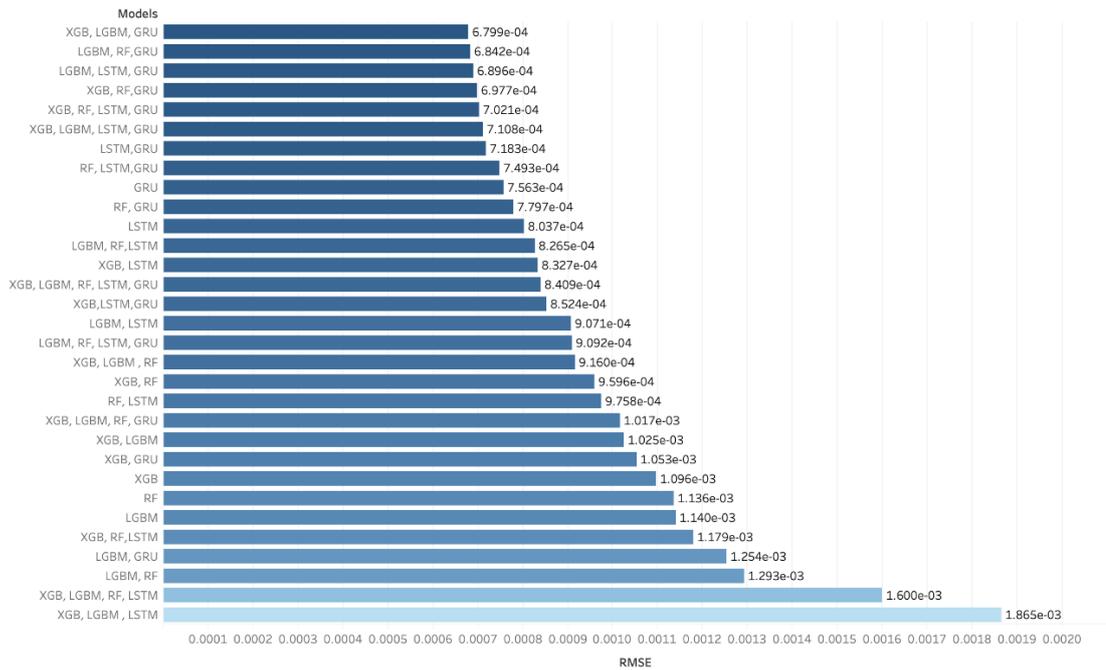

**Figure 20**: *The RMSE of different models' combinations on AUD_USD*

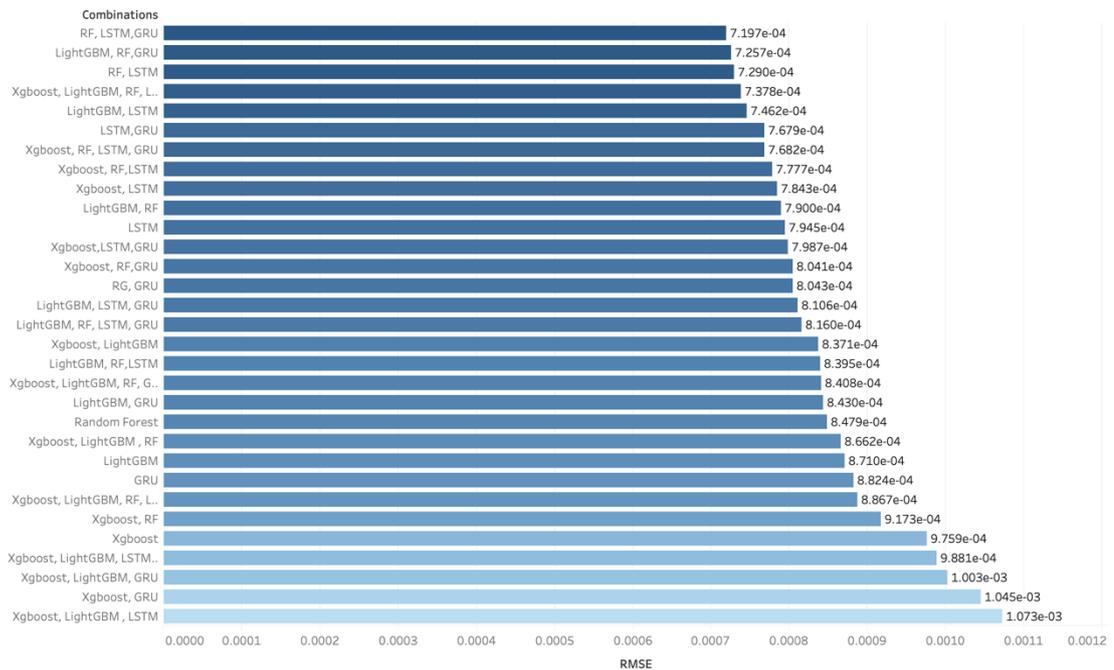

**Figure 21**: *The RMSE of different models' combinations on EUR_USD*

The performances of 31 stacking combinations are shown in Figure 21. The stacking system using Random Forest, LSTM, and GRU has the smallest RMSE overall. The RMSE of this stacking system is 7.2E-04, which is much better than RMSE 7.94E-04 of LSTM, the single best one. Hence, in the dataset of M15 EUR-USD, we finally choose to use Random Forest, LSTM, and GRU in the first layer of the stacking system.

The result shows that the performance is improved with only certain combinations, and overall, the combinations with both LSTM and GRU outperforms other models.

## 6. DISCUSSION

The forex trading market is a global speculative market. By far, it is the largest market in terms of quantitative trading. The characteristics of forex trading distinguish it from other quantitative trading markets. Unlike we buy stocks with money in the stock market, what is exchanged in forex trading is the "money" itself, and the main action is to exchange one currency to another. With its international transparency, maturity, and standardization, it has attracted the extensive participation of banks, large enterprises, funds, and investment institutions as well as individual investors from all over the world.

The primary goal of this project is to predict the highest high price of the next 5 timestamps based on the historical data with machine intelligent models. We have proposed a new feature selection approach to filter the most important features for a deep learning model and a stacking model to combine and improve the performance from multiple models. Both approaches significantly increased the performance.

The raw datasets are the OHLC data of a certain period downloaded from OANDA without any other features. For feature engineering, we first calculated 139 features including overlap studies indicators, momentum indicators, volatility indicators, price transform, cycle indicators and pattern recognition functions.

We also used the ARIMA predictions for the close and high price of next timestamp as features. ARIMA by itself is already a mature statistical model, as AIC function returns a q value of 0 which means we only used the AR method. So, we did not use ARIMA as a model as it provides a lagging prediction. Instead, we keep the ARIMA prediction as another feature. The feature importance recap result also proved that the ARIMA prediction is vital for highest high price prediction.

We thought that since all these features are calculated based on the OHLC value, they are all related to our target. Besides, the deep learning models already have built-in mechanisms to filter the features, so we did not apply any feature selection approach. With all these features, we applied many deep learning models including LSTM, CNN, GRU, Transformer with LSTM, Reinforcement learning. However, these models did not perform well on our dataset. Some model did not even converge in experiments. We revise the processes and the model. By analyzing the features, we found that over 70 columns of cycle indicators and pattern recognition functions are 0s. These values would make the data noisy and mislead the prediction.

Then we drop these features and retrain the model. With LSTM we got the lowest RMSE on next timestamp close price prediction of 0.006. Although it is still high, it is much better than our previous results.

From this experiment, we found that our dataset is too noisy, and we are not able to get accurate results with such a combination of dataset and model. Therefore, we decided to apply denoise and feature selection approaches to improve the performance.

For data denoise, we first did research about wavelet denoise. Most of the experiments shows that it would improve the model performance. However, [21] proved that wavelet denoise would lead to data leakage, it may use future data to denoise historical data. So, we gave up wavelet denoise and turn to find a new way for feature selection.

In previous study, we have learned the feature importance calculation in tree-based models. Consequently, we did research about feature importance calculation in deep learning models and tried to implement it in our experiments. We found that the SHAP package based on the game theory could explain the importance of each feature in the dataset. However, it has a high computational cost as it tries all combinations among the features. With financial data, we could easily get hundreds of features by changing the time which means that SHAP is not a valid approach. At last, we decided to find an approach to combine the results from tree-based models and deep learning models.

In general, we cannot apply the feature importance from on model in another as the training process is different, which means we cannot directly apply the result from tree-based model in deep learning model. Our experiments also proved that applying the feature importance on other models would increase the performance, but it is not the most suitable feature set.

To allow the feature importance score from the tree-based model carries the performance of the selected features in deep learning model, we feed the selected feature in deep learning and divide the feature importance score by the RMSE then sum the importance score of each feature. The final importance will contain information from both tree-based feature importance and deep learning model performance. Compare with SHAP, this approach is much faster. We tested this approach on AUD_USD and EUR_USD with both LSTM and GRU. The result shows that the RMSE is significantly dropped.

Moreover, we found that the tree-based model also could also perform well on the datasets, so we decided to implement a stacking model to combine the results from five models: XGBoost, Random Forest, LightGBM, LSTM and GRU. We tried all possible combinations, and the result is further improved.

There were four main challenges during our research and experiments:

- Lack of domain knowledge

  All group members have no financial trading experience. We had to spend time to learn the financial fundamentals regarding this project such as what is forex trading and its characteristics, technical indicators, why we need to predict several timestamps ahead rather than predict next one timestamp.

- Lack of timeseries data forecasting experience

  All group members are major in data science but all of us do not have timeseries data forecasting experience. In previous studies, we have learned several machine learning models such as XGB, K-NN and some basic RNN, but each row in the raw dataset is not related. For financial data, the previous data will impact the future data and we had to learn how to modify the dataset so that they are in a sequence while feeding into the models.

- Hardware infrastructure limitation

  Due to the COVID-19, we cannot go to the campus and use advanced infrastructures in the labs. Most group members can only use laptops for model training. The limitation of computational resource makes us can only experiment with small datasets and cannot apply grid search on hyper parameters.

- Different levels of coding experience within the team

All the codes of this project are in python. Some group members do not have enough coding experience and had to spend time on practicing.

What we learned from this project:

- Always analyze the data before building model

    In the first few weeks, we spend lots of time on implementing different deep learning models. However, the performance is very bad as the data is too noisy after we added cycle indicators and pattern recognition functions as features. Most values of these two types of indicators are 0s and 1s, if we made detailed data analysis, we might save these times.

- Future selection is important

    After we feed the selected features using tree-based models in deep learning models, the performance is better than feeding all features. The results from our feature selection recap approach also proved that Propper feature selection would benefit the model.

- Group work can improve work efficiency

    Group members have different knowledge backgrounds and experience, and their respective areas of expertise are also different. We divided the tasks for each member so that everyone can contribute to the project. As part of the goal of this project is to improve our understanding on AI techniques, we also helped each other in coding.

## 7. LIMITATIONS AND FUTURE WORKS

While the result to date shows that our approach significantly improved the performance of highest high price prediction, there are much more we would like to achieve. The limitation of our work is largely due to the time constraint. The project needs to be done in 13 weeks. In the first few weeks, we spend lots of time on testing deep learning models but did not get any acceptable results. The feature importance recap and model stacking approaches was proposed and implemented in week 10. Therefore, we did not have enough time to test more deep learning models and currency pairs.

We have tested our approach with two currency pairs, two deep learning models and stacked these two model with three tree-based models. The results proved that our approach would improve the RMSE but to get more stable results, we need to test with more currency pairs and models.

Moreover, this project did not involve any backtesting to test if the trading strategy based on this prediction would make profit. This is crucial for financial data forecasting as the purpose of price forecasting is to help the investors to make money. The prediction with low RMSE is not guaranteed high profit margin. The prediction accuracy is evaluated with RMSE, despite it predicted the price higher or lower, the RMSE may stays the same. However, in financial trading, mis-predicting the increase or decrease may lead to profit loss. Therefore, to prove our approach would benefit the investors, we should apply back testing with some pre-defined trading strategies.

Since there are overlapping among the importance features in different models, we could run the experiments more times to test which feature is always important and try to predict only with these features. We can generate a solid feature set for forex forecasting if possible. Then, when other researchers want to test a model, they can simply feed these features.

Alternatively, we can also make predictions about the lowest low prices of following timestamps in the future. Then design trading strategy with both highest high and lowest low prediction. We can also compare the selected features from these two predictions and analyse why such feature is importance from financial aspects. Furthermore, we can extend this prediction to OHLC data, and some technical indicators then design trading strategies with a predicted dataset and apply backtesting to test the profitability.

For building trading strategies, rather than using rule-based trading, we can apply reinforcement learning to further improve the strategy. Reinforcement learning will reward the correct actions and punish actions that lost money. We can compare the performance between the strategy from reinforcement learning and from expertise rules.

# 8. APPENDIX

| ID | RMSE | MAE | Combinations |
|---|---|---|---|
| 0 | 1.10E-03 | 6.22E-04 | Xgboost |
| 1 | 1.14E-03 | 6.23E-04 | LightGBM |
| 2 | 1.14E-03 | 6.42E-04 | Random Forest |
| 3 | 8.04E-04 | 6.49E-04 | LSTM |
| 4 | 7.56E-04 | 5.31E-04 | GRU |
| 5 | 1.03E-03 | 9.02E-04 | Xgboost, LightGBM |
| 6 | 9.60E-04 | 8.08E-04 | Xgboost, RF |
| 7 | 8.33E-04 | 5.88E-04 | Xgboost, LSTM |
| 8 | 1.05E-03 | 7.96E-04 | Xgboost, GRU |
| 9 | 1.29E-03 | 1.16E-03 | LightGBM, RF |
| 10 | 9.07E-04 | 7.77E-04 | LightGBM, LSTM |
| 11 | 1.25E-03 | 9.99E-04 | LightGBM, GRU |
| 12 | 9.76E-04 | 8.46E-04 | RF, LSTM |
| 13 | 7.80E-04 | 5.41E-04 | RG, GRU |
| 14 | 7.18E-04 | 4.98E-04 | LSTM,GRU |
| 15 | 9.16E-04 | 7.88E-04 | Xgboost, LightGBM , RF |
| 16 | 1.87E-03 | 1.64E-03 | Xgboost, LightGBM , LSTM |
| 17 | 6.80E-04 | 5.02E-04 | Xgboost, LightGBM, GRU |
| 18 | 1.18E-03 | 9.42E-04 | Xgboost, RF,LSTM |
| 19 | 6.98E-04 | 5.39E-04 | Xgboost, RF,GRU |
| 20 | 8.52E-04 | 5.89E-04 | Xgboost,LSTM,GRU |
| 21 | 8.27E-04 | 6.99E-04 | LightGBM, RF,LSTM |
| 22 | 6.84E-04 | 4.98E-04 | LightGBM, RF,GRU |
| 23 | 6.90E-04 | 5.31E-04 | LightGBM, LSTM, GRU |
| 24 | 7.49E-04 | 5.08E-04 | RF, LSTM,GRU |
| 25 | 1.60E-03 | 1.49E-03 | Xgboost, LightGBM, RF, LSTM |
| 26 | 1.02E-03 | 8.90E-04 | Xgboost, LightGBM, RF, GRU |
| 27 | 7.11E-04 | 5.61E-04 | Xgboost, LightGBM, LSTM, GRU |
| 28 | 7.02E-04 | 5.51E-04 | Xgboost, RF, LSTM, GRU |
| 29 | 9.09E-04 | 6.56E-04 | LightGBM, RF, LSTM, GRU |
| 30 | 8.41E-04 | 5.90E-04 | Xgboost, LightGBM, RF, LSTM, GRU |

**Figure 22**: *Stacking performance of 31 combinations on AUD_USD*

| ID | RMSE | MAE | Combinations |
| --- | --- | --- | --- |
| 0 | 9.76E-04 | 7.16E-04 | Xgboost |
| 1 | 8.71E-04 | 6.29E-04 | LightGBM |
| 2 | 8.48E-04 | 6.16E-04 | Random Forest |
| 3 | 7.94E-04 | 5.45E-04 | LSTM |
| 4 | 8.82E-04 | 7.14E-04 | GRU |
| 5 | 8.37E-04 | 6.36E-04 | Xgboost, LightGBM |
| 6 | 9.17E-04 | 7.08E-04 | Xgboost, RF |
| 7 | 7.84E-04 | 5.67E-04 | Xgboost, LSTM |
| 8 | 1.04E-03 | 7.57E-04 | Xgboost, GRU |
| 9 | 7.90E-04 | 6.17E-04 | LightGBM, RF |
| 10 | 7.46E-04 | 5.07E-04 | LightGBM, LSTM |
| 11 | 8.43E-04 | 5.78E-04 | LightGBM, GRU |
| 12 | 7.29E-04 | 5.53E-04 | RF, LSTM |
| 13 | 8.04E-04 | 5.39E-04 | RG, GRU |
| 14 | 7.68E-04 | 6.11E-04 | LSTM,GRU |
| 15 | 8.66E-04 | 6.92E-04 | Xgboost, LightGBM , RF |
| 16 | 1.07E-03 | 8.07E-04 | Xgboost, LightGBM , LSTM |
| 17 | 1.00E-03 | 7.86E-04 | Xgboost, LightGBM, GRU |
| 18 | 7.78E-04 | 5.40E-04 | Xgboost, RF,LSTM |
| 19 | 8.04E-04 | 5.80E-04 | Xgboost, RF,GRU |
| 20 | 7.99E-04 | 6.20E-04 | Xgboost,LSTM,GRU |
| 21 | 8.40E-04 | 6.94E-04 | LightGBM, RF,LSTM |
| 22 | 7.26E-04 | 5.51E-04 | LightGBM, RF,GRU |
| 23 | 8.11E-04 | 5.71E-04 | LightGBM, LSTM, GRU |
| 24 | **7.20E-04** | **4.83E-04** | **RF, LSTM,GRU** |
| 25 | 8.87E-04 | 6.24E-04 | Xgboost, LightGBM, RF, LSTM |
| 26 | 8.41E-04 | 6.08E-04 | Xgboost, LightGBM, RF, GRU |
| 27 | 9.88E-04 | 8.30E-04 | Xgboost, LightGBM, LSTM, GRU |
| 28 | 7.68E-04 | 5.42E-04 | Xgboost, RF, LSTM, GRU |
| 29 | 8.16E-04 | 5.52E-04 | LightGBM, RF, LSTM, GRU |
| 30 | 7.38E-04 | 5.32E-04 | Xgboost, LightGBM, RF, LSTM, GRU |

**Figure 23**: *Stacking performance of 31 combinations on EUR_USD*